\documentclass{iaus}
\usepackage{graphicx}

\title[Modeling GRB Host Galaxies] 
{Simulating High-Redshift Disk Galaxies:\\ Applications to Long Duration \\ Gamma-Ray Burst Hosts}

\author[Brant E. Robertson]   
{Brant E. Robertson$^{1,2,3}$}

\affiliation{$^1$Kavli Institute for Cosmological Physics, and
\\ Department of Astronomy and Astrophysics, University of Chicago,
\\ 933 East 56th Street, Chicago, IL 60637, USA
\\ $^2$Enrico Fermi Institute, 5640 South Ellis Avenue, Chicago, IL 60637, USA
\\ $^3$Spitzer Fellow\\ email: {\tt brant@kicp.uchicago.edu}}

\pubyear{2008}
\volume{254}  
\pagerange{--}
\jname{The Galaxy Disk in Cosmological Context}
\editors{J. Anderson, J. Bland-Hawthorn \& B. Nordstrom, eds.}
\begin{document}

\maketitle

\begin{abstract}
The efficiency of star formation governs many observable properties of 
the cosmological galaxy population, yet many current models of 
galaxy formation largely ignore the important physics of star formation 
and the interstellar medium (ISM).
Using hydrodynamical simulations of disk galaxies
that include a treatment of the molecular ISM and star formation in
molecular clouds (Robertson \& Kravtsov 2008), we study the influence
of star formation efficiency and molecular hydrogen abundance on the
properties of high-redshift galaxy populations.  In this work, we 
focus
on a model of low-mass, star forming galaxies at 
$1\lesssim z\lesssim2$ that may host long duration gamma-ray bursts (GRBs).
Observations of GRB hosts have revealed a population of faint systems with
star formation properties that often differ from Lyman-break 
galaxies (LBGs) and
more luminous high-redshift field galaxies.  Observed GRB
sightlines are deficient in molecular hydrogen, but
it is unclear to what degree this deficiency owes to intrinsic properties
of the galaxy or the impact the GRB has on its environment.
We find that hydrodynamical simulations of low-stellar mass systems at
high-redshifts
can reproduce the observed star formation rates
and efficiencies of 
GRB host galaxies at redshifts
$1\lesssim z \lesssim2$. 
We show that the compact structure of low-mass high-redshift GRB hosts
may lead to a molecular ISM fraction of a few tenths, 
well above that observed in individual GRB sightlines.  
However, the star formation rates of observed GRB host galaxies 
imply molecular gas masses of $10^{8}-10^{9}~M_{\odot}$ similar to
those produced in the simulations, and may therefore imply fairly large
average H$_{2}$ fractions in their ISM.
\keywords{Galaxies:high-redshift, galaxies:ISM, gamma rays: bursts}
\end{abstract}

\firstsection 
\section{Introduction}

To improve the physical description of star formation in
hydrodynamical simulations of galaxies, 
\cite[Robertson \& Kravtsov (2008)]{robertson2008a}
implemented a new model for the ISM that includes 
low-temperature ($T<10^{4}$K) cooling, directly ties the 
star formation rate to the molecular gas density, 
and accounts for the destruction of molecular hydrogen by an 
interstellar radiation field (ISRF) from young stars.
They used simulations to study the relation between star formation 
and the ISM in galaxies and demonstrated that, for the first time, 
their new model simultaneously reproduces the molecular gas and 
total gas Kennicutt-Schmidt (KS) relations, the connection between 
star formation and disk rotation, and the relation between 
interstellar pressure and the fraction of gas in molecular form
\cite[(e.g. Wong \& Blitz 2002, Blitz \& Rosolowsky 2006)]{wong2002a,blitz2006a}.  
The capability of this model to reproduce both the
star formation efficiency and molecular abundance of nearby systems
makes it useful for simulating low-mass galaxies that have
suppressed H$_{2}$ abundances (and whose star formation
rates would be overestimated in common treatments of star formation 
based on the KS relation) and high-redshift galaxies
whose structural properties may vary substantially from local
systems (and may therefore not have the same KS
relation normalization).  The model should be especially useful
for studying low-mass galaxies at high-redshift,
such as long duration gamma-ray burst (GRB) host galaxies at 
$1\lesssim z \lesssim2$, which is the focus of this work.

The highly-energetic phenomena known as GRBs
were discovered over forty years ago \cite[(Klebesadel et al. 1973)]{klebesadel1973a},
but their extragalactic origin was confirmed only in the last decade 
\cite[(e.g., Metzger et al. 1997)]{metzger1997a}.  Since then, the properties
of the cosmological population of galaxies that host GRBs have been
increasingly well-studied \cite[(e.g., Bloom et al. 2002, Le Floc'h et al. 2006, Prochaska et al. 2006, Berger et al. 2007a,b)]{bloom2002a,le_floch2006a,prochaska2006a,berger2007a,berger2007b}.
Recently, interest in long duration GRB galaxy hosts as possible tracers 
of the
global star formation history of the universe has motivated systematic
studies of their star formation efficiencies and stellar masses
\cite[(Castro Cer\'on et al. 2008, Savaglio et al. 2008)]{castro_ceron2008a,savaglio2008a}.  
These studies have found that high-redshift GRB hosts have 
small stellar masses ($\log M_{\star}\sim 9.3$) and moderate star formation
rates ($\mathrm{SFR}\sim2.5~M_{\odot}~\mathrm{yr}^{-1}$).  Compared with
other high-redshift galaxy populations, GRB hosts tend to have lower star
formation rates at fixed stellar mass compared with Lyman-break galaxies
and lower stellar masses at fixed star formation rate compared with field
galaxies \cite[(for details, see Savaglio et al. 2008)]{savaglio2008a}.

Spectroscopic studies of GRB sightlines have provided additional information about
the post-explosion character of the host galaxy ISM.  \cite[Tumlinson et al. (2007)]{tumlinson2007a}
failed to detect H$_{2}$ in five GRB sightlines and suggested that low metallicity
and large far ultraviolet ISRF strengths ($10-100\times$ the Milky Way value) were
responsible for destroying molecular hydrogen in GRB hosts.  They interpreted the
lack of vibrationally excited H$_{2}$ lines as evidence against the GRB destroying
its parent molecular cloud, but noted various caveats to this conclusion such as
the parent cloud size or cloud photodissociation before to the GRB.  
\cite[Whalen et al. (2008)]{whalen2008a} used one-dimensional radiative hydrodynamical
calculations to show that GRBs can ionize nearby neutral hydrogen, but suggested
that an additional
ISRF
is necessary to remove molecular hydrogen from the nearby ISM.  
\cite[Prochaska et al. (2008)]{prochaska2008a} studied NV absorption in GRB sightlines,
and argued that if nitrogen ionization by GRB afterglows leads to NV absorption then
the observations support a scenario where dense, molecular cloud-like environments serve
as the sites of GRBs.

Given the increasingly detailed studies of GRB hosts, their interesting ISM and star formation
properties, and their low stellar masses, a theoretical study of GRB host galaxy
analogues using hydrodynamical simulations that include a treatment of the molecular ISM
is warranted.  Below, we present simulations of a model GRB host galaxy that include a
prescription for the molecular ISM and star formation in molecular clouds
\cite[(Robertson \& Kravtsov 2008)]{robertson2008a}.  We use the simulations to examine
the star formation efficiency and molecular hydrogen content of galaxies with structural
properties similar to those expected for low-mass galaxies at $1\lesssim z \lesssim 2$.
Below, we discuss our methodology and present some initial results.

\section{Methodology}

To study the properties of long duration GRB host galaxies, we simulate a numerical
model of an isolated galaxy using a version of the N-body/Smoothed Particle Hydrodynamics code
GADGET \cite[(Springel et al. 2001, Springel 2005b)]{springel2001a,springel2005b}
that incorporates a model for the molecular ISM \cite[(Robertson \& Kravtsov 2008)]{robertson2008a}.
For details regarding the numerical galaxy models, simulation methodology, and ISM model, we 
refer the reader to \cite[Springel et al. (2005a)]{springel2005a}, 
\cite[Robertson et al. (2006a,b)]{robertson2006a,robertson2006b}, and
\cite[Robertson \& Kravtsov (2008)]{robertson2008a}, but a brief summary follows.

The numerical galaxy model is designed to approximate the properties of 
$1\lesssim z\lesssim2$ GRB host
galaxies as determined by \cite[Savaglio et al. 2008]{savaglio2008a}.  
The stellar disk
mass of the system is set to $\log M_{\star} = 9.3$, with a gas fraction 
of $f_{\mathrm{gas}}=0.5$ \cite[(appropriate for high-redshift, see Erb et al. 2006)]{erb2006a},
which implies a total virial mass of 
$\log M_{\mathrm{vir}}=10.9$ for a typical disk baryon fraction
of $f_{\mathrm{b}}=0.05$.  The virial radius is set appropriately for a 
halo with virial
mass $M_{\mathrm{vir}}$ at $z\sim2$.  The exponential disk scale length 
was fixed
according to the \cite[Mo et al. (1998)]{mo1998a} formalism, including 
the adjustment for an
effective \cite[Navarro et al. (1996)]{navarro1996a} dark matter halo 
concentration of 
$c_{\mathrm{NFW}}=6$
\cite[(also appropriate for the chosen virial mass and redshift, see 
Bullock et al. 2001)]{bullock2001a}
and a spin of $\lambda=0.05$.
The density field of the dark matter halo follows the 
\cite[Hernquist (1990)]{hernquist1990a} profile, 
while 
the velocity fields of the dark matter halo and the exponential stellar 
disk are set using 
the \cite[Hernquist (1990)]{hernquist1990a} distribution function and
the epicyclical approximation, respectively.
The numerical realizations of the stellar disk, gaseous disk, and dark 
matter halo are initialized
with 
$N_{\mathrm{disk},\star}=4\times10^{5}$, 
$N_{\mathrm{disk},\mathrm{gas}}=4\times10^{5}$, and 
$N_{\mathrm{DM}}=4\times10^{5}$
particles, and are evolved with a gravitational softening of 
$\epsilon = 70~\mathrm{pc}$.
The simulation is calculated for a duration of $t\sim1~\mathrm{Gyr}$, or 
about the
time between redshift $z\sim2$ and $z\sim1.5$.

\begin{figure}[t]
\begin{center}
 \includegraphics[width=5.0in]{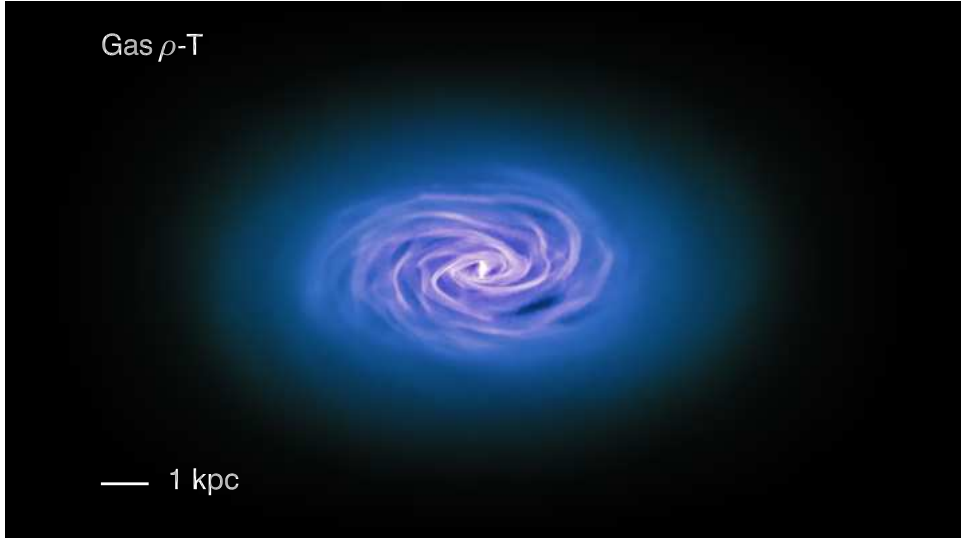} 
 \caption{Simulated long duration Gamma-Ray Burst (GRB) host galaxy analogue at $z\sim1.5$.  Shown is the gas surface 
density of the GRB host (image intensity), color coded by the median interstellar medium
temperature (purple 
regions have $T<10^{3}$K, while blue regions have $T\gtrsim10^{4}$K).  The simulated galaxy has a
stellar mass $\log M_{\star} \approx 9.3$ and a star formation rate $\mathrm{SFR}\approx1.2 M_{\odot} \mathrm{yr}^{-1}$, 
similar to
high-redshift GRB host galaxies 
\cite[(e.g., Castro Cer\'on et al. 2008, Savaglio et al. 2008)]{castro_ceron2008a,savaglio2008a}.  The simulations
include the \cite[Robertson \& Kravtsov (2008)]{robertson2008a} model of the molecular ISM, enabling a
study of the connection between star formation rate, galaxy properties, and H$_{2}$ abundance in GRB
hosts.}.
\label{fig1}
\end{center}
\end{figure}

\begin{figure}[t]
\begin{center}
 \includegraphics[width=3.4in]{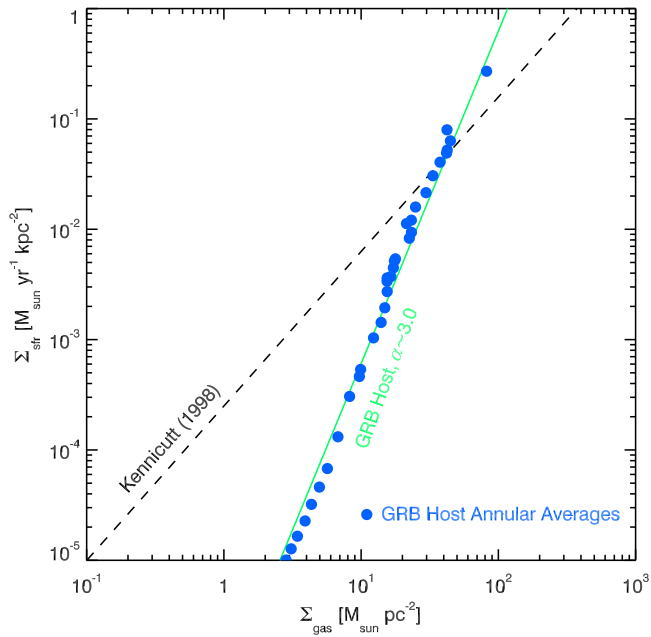} 
 \caption{Kennicutt-Schmidt relation for a simulated GRB host galaxy at $z\sim1.5$.  
Shown is the star formation rate surface density $\Sigma_{\mathrm{SFR}}$ as a function of total gas 
surface density $\Sigma_{\mathrm{gas}}$, measured in annuli (blue dots).  The average 
Kennicutt-Schmidt relation of the GRB host has a
steeper power-law index ($\alpha\sim3.0$, green line) than the disk-averaged relation measured by 
Kennicutt (1998; $\alpha\sim1.4$, dashed line), owing to the 
suppression of H$_{2}$ in the galaxy exterior by the interstellar radiation field and the low 
ISM metallicity
\cite[(for a detailed discussion, see Robertson \& Kravtsov 2008)]{robertson2008a}.}
\label{fig2}
\end{center}
\end{figure}

The simulation includes a treatment of the physics of the ISM and star 
formation following the model 
presented by \cite[Robertson \& Kravtsov (2008)]{robertson2008a}, and 
interested readers
should examine that work for details.  The photoionization code CLOUDY 
\cite[(Ferland et al. 1998)]{ferland1998a}
is used to tabulate the cooling rate, heating rate, molecular abundance, 
and related properties of
gas as a function of density, temperature, metallicity, and local 
interstellar radiation field (ISRF) 
strength.
The star formation rate is calculated by converting the 
\it molecular \rm gas density to stars on a 
timescale
that scales with the local dynamical time, with an efficiency set to match the star formation efficiency
per free fall time in local molecular clouds 
\cite[(e.g., Krumholz \& McKee 2005, Krumholz \& Tan 2007)]{krumholz2005a,krumholz2007a}.
The local ISRF spectral shape is fixed to the local Milky Way ISRF 
inferred by 
\cite[Mathis et al. (1983)]{mathis1983a}, but
the ISRF strength scales with the local star formation rate 
density (i.e., young, massive stars supply the local 
ultraviolet
radiation field).  The abundance of molecular gas 
tracked using CLOUDY includes the photodissociative and
heating effects of this ISRF, and thereby includes a 
coarse accounting of the regulatory impact of the 
ISRF  
on star formation in molecular clouds.

\section{Results}

Figure \ref{fig1} shows the gaseous structure of 
the GRB host galaxy model at $z\sim1.5$ (after $900~\mathrm{Myr}$ of 
evolution).  The figure shows the gas surface
density of the system (image intensity) and the median temperature of the 
local ISM (purple regions have temperatures $T\lesssim10^{3}$K, 
while blue regions have $T\gtrsim10^{4}$K).  The system has a rotational
velocity of $v_{\mathrm{rot}}\approx100~\mathrm{km}~\mathrm{s}^{-1}$
and a disk scale length of $R_{\mathrm{d}}\approx1.5~\mathrm{kpc}$.
During the simulation the star formation rate of the system varies in the
range $\mathrm{SFR}\approx0.5-2.5~M_{\odot}~\mathrm{yr}^{-1}$,
while the specific star formation rate is
$\mathrm{SFR}/M_{\star} \approx 0.17-1.1~\mathrm{Gyr}^{-1}$.  These
properties are consistent with the properties of high-redshift GRB host galaxies
determined by \cite[Savaglio et al. (2008)]{savaglio2008a}, who find
star formation rates of $\mathrm{SFR}\sim0.1-10$ and specific star
formation rates of $M_{\star}/\mathrm{SFR}\sim0.1-10~\mathrm{Gyr}$.

The compactness of the system leads to a dense ISM and a considerable molecular gas
fraction.
The global, mass-weighted molecular abundance declines from 
$f_{\mathrm{H}2}\sim 0.5$ at $z\sim2$ to $f_{\mathrm{H}2}\sim0.3$ at $z\sim1.5$.
As a function of radius, the molecular fraction declines from $f_{\mathrm{H}2}\sim1$
near the center of the galaxy to $f_{\mathrm{H}2}\sim0.1-0.3$ beyond a disk scale
radius.  The typical star formation rate-weighted radius of the system is 
$r_{\mathrm{SFR}}\sim0.8~\mathrm{kpc}$, where the ISM molecular fraction is
$f_{\mathrm{H}2}\sim0.6$.  Hence, if observed high-redshift GRB host galaxies are similar
in nature to this simulated system, GRBs will likely occur in molecular-rich
regions.
While these molecular abundances are consistent with the spectroscopic studies by 
\cite[Prochaska et al. (2008)]{prochaska2008a}, they are well above 
the observed H$_{2}$ abundance along GRB sightlines 
\cite[(e.g., Tumlinson et al. 2007)]{tumlinson2007a}.  In this model, the compact
and dense structure of the high-redshift GRB host prevents the diffuse
ISRF from suppress the H$_{2}$ to levels observed in GRB sightlines.  If an ISRF
is responsible for suppressing H$_{2}$ to observed levels in GRB sightlines, 
it may be generated by discrete point sources nearby the GRB in a manner not
captured by the diffuse ISRF included in these simulations.

We note that in order to supply the star formation efficiency for GRB hosts determined
by \cite[Savaglio et al. (2008)]{savaglio2008a}, GRB hosts may need to be
fairly molecule rich if their structure is similar to the simulated high-redshift
galaxy analogues presented here.
Figure \ref{fig2} shows the total gas Kennicutt-Schmidt (KS) relation for the GRB host galaxy
analogue, measured in annuli with a width of $\Delta r = 100~\mathrm{pc}$.  Plotted is
the star formation rate density $\Sigma_{\mathrm{SFR}}$ as a function of the total gas
surface density $\Sigma_{\mathrm{gas}}$ (blue points), compared with the mean disk-averaged
trend determined by \cite[Kennicutt (1998, dashed-line)]{kennicutt1998a}.  
The central concentration of
molecular gas causes the total gas KS relation of the simulated GRB host
galaxy analogue to be steeper than the disk-averaged relation.  In order to supply the
observed star formation rate of $\mathrm{SFR}\sim1-10~M_{\odot} \mathrm{yr}^{-1}$, as this
simulated galaxy does, the typical consumption timescales of $\sim 100~\mathrm{Myr}$ for
molecular gas imply a reservoir of roughly $M_{\mathrm{H}2}\sim0.1-1\times10^{9}~M_{\odot}$
(the simulated system has $M_{\mathrm{H}2}\sim2-7\times10^{8}~M_{\odot}$ during its
evolution).
Since observed GRB hosts have stellar masses of only $\log M_{\star}\sim9.3$ \cite[(Savaglio et al. 2008)]{savaglio2008a}, 
the inferred molecular fraction
of the ISM should be large even for very gas rich systems.

Overall, we find that under standard assumptions about the mass and redshift scalings
of galaxy structure hydrodynamical simulations of disk galaxies with stellar masses of
$\log M_{\star}\sim9$ that utilize a model for the molecular ISM and star formation
in molecular clouds \cite[(Robertson \& Kravtsov 2008)]{robertson2008a} can reproduce
the observed star formation rates and efficiencies of GRB hosts 
\cite[(e.g., Castro Cer\'on et al. 2008, Savaglio et al. 2008)]{castro_ceron2008a,savaglio2008a}.
The star formation in both observed GRB hosts and the simulated GRB host analogue
presented here is efficient for their low stellar masses, and to supply the observed
range of star formation rates ($\mathrm{SFR}\sim0.1-10~M_{\odot}~\mathrm{yr}^{-1}$) the
molecular gas content such systems may need to be considerable ($f_{\mathrm{H}2}\gtrsim0.1$).
While this result is consistent with observations that suggest GRBs occur in dense,
potentially molecular-rich
regions of the ISM \cite[(e.g. Prochaska et al. 2008)]{prochaska2008a}, more work is
needed to reconcile such results with the low molecular abundance observed in GRB sightlines
\cite[(e.g. Tumlinson et al. 2007)]{tumlinson2007a} if GRBs cannot efficiently destroy 
H$_{2}$ in the ISM \cite[(Whalen et al. 2008)]{whalen2008a}.


\end{document}